\documentclass[11pt,twoside]{article}
\usepackage{asp2010}

\resetcounters

\begin{document}

\title{Morphological classification of post-AGB stars}
\author{A. Manchado$^{1,2}$, D. A. Garc\'{\i}a-Hern\'andez $^1$, E. Villaver$^3$, J. Guironnet de Massas$^4$}
\affil{$^1$Instituto de Astrof\'{\i}sica de Canarias, C/ Via L\'actea
s/n, 38200 La Laguna, Spain; agarcia@iac.es, amt@iac.es}
\affil{$^2$ Consejo Superior de Investigaciones Cient\'{\i}ficas, Spain}
\affil{$^3$ Departamento de F\'{\i}sica Te\'orica C-XI, Universidad
Aut\'onoma de Madrid, E-28049 Madrid, Spain; eva.villaver@uam.es}
\affil{$^4$ University Joseph Fourier, Grenoble University Joseph Fourier, 3840
Grenoble, France}

\begin{abstract}
We present a complete study of the morphology of post-Asymptotic Giant
Branch
(AGB) stars. Post-AGB is a very short evolutionary phase between the end
of the
AGB and the beginning of the Planetary Nebula (PN) stage (between 100
and 10,000
yrs). We have defined the end of the post-AGB phase and the beginning of
the PN
phase when the star is hot enough to fully ionize the hydrogen envelope.
Post-AGB stars have a circumstellar shell that is illuminated by the
central
stars or partially ionized. However, this circumstellar shell is too
small to be resolved
 from ground-based observations. Thus, we have
used data from the Hubble Space Telescope (HST) database to resolve
these
shells. About 150 post-AGB were found in this database. Here we present
the
preliminary results on their morphological classification and the
correlation
with several parameters such as galactic latitude and IRAS fluxes. Our
preliminary results show that 40 \% of the sample are stellar-like (S),
33 \%
bipolar (B), 12 \% multi-polar (M) and 15 \% elliptical (E).
\end {abstract}


As has been shown over the past 30 years, the study of the Planetary
Nebulae
(PNe) morphologies provides valuable information about the late stages
of
stellar evolution.  The processes that lead to the different
morphologies
observed in PNe (Manchado 2003) are not known in detail. 

Because of the short-lived (~10$^2$ - 10$^4$ yrs) nature of the post-AGB
transition phase,
the number of known objects is scarce. In addition, the physical size of
the observed nebulosities is very small (between 1 to 10 arcseconds on
the
sky), making ground-based observations very difficult. The post-AGB
nebulosities can display an incipient bipolar structure at a very early
stage in
the post-AGB phase, as has been demonstrated by the Hubble Space
Telescope (HST)
observations of different samples of post-AGB stars (e.g., Ueta et al.
2000,
Sahai et al. 2007, Siodmiak et al 2008). However, there is a general
perception
in the literature that most of the observed post-AGB stars display
asymmetrical
(e.g., bipolar, multi-polar, etc.) structures. This is in an apparent
contradiction with the morphologies observed in more complete samples of
evolved
PNe, where round (25\%), elliptical (58\%) and bipolar and multi-polar
(17\%) PNe are
observed (Manchado 2003). This contradiction is most likely related to a
strong bias when selecting post-AGB samples, as well as the fact that
almost all
authors in the literature report and discuss the high-spatial resolution
images only of the post-AGB objects that show some extended structure.
Here we
present a complete and less biased study of the morphology of post-AGB
stars in
order to shed some light on the possible reasons of this apparent
contradiction between the morphologies observed in PNe and their
immediate
precursors, the post-AGB stars.

We have selected a large sample of post-AGB candidates showing infrared
colors
similar to PNe (Manchado et al. 1989, Garcia-Lario et al. 1997). This
sample was
complemented with the Torun Catalogue of post-AGB stars and related
objects
(Szczerba et al. 2007) as well as with previous surveys based on
ground-based
optical spectroscopy (Suarez et al. 2006) and HST imaging of post-AGB
stars
(e.g., Ueta et al. 2000, Sahai et al. 2007, Siodmiak et al 2008). Note
that we
have defined the end of the post-AGB phase and the beginning of the PN
phase
when the star is hot enough (spectral type earlier than B) to fully
ionize the
hydrogen envelope.

We have searched the HST database (http://archive.stsci.edu/) and
downloaded
images for 140 post-AGB stars obtained with the ACS, WFPC, WFPC2, and
NICMOS
instruments onboard HST. The available images were analyzed and all
post-AGB
stars were morphologically classified as stellar-like, round-elliptical,
bipolar
and multi-polar.

From the sample of 140 post-AGB star, 25 could not classified, because
several factors,
e.g. we could not identify the star or the morphology did not follow the
classification
scheme. The rest of the sample could be divided in four morphological
classes, with 46 (40\%)objects
being stellar like (S), 38 (33\%) Bipolar (B), 14 (12\%) multi-polar (M)
and 17 (15\%) elliptical (E).
It is specially remarkable that non round post-AGB were found, on
contrary which is found in PNe(25\%) 

\section{Results}

When studying, the Galactic distribution of post-AGB and
PNe, it is remarkable that most of the post-AGB are concentrated around the
galactic plane in comparison with PNe.
S type seems to be the around the galactic bulge, and have a higher
Galactic latitude distribution together
with the E type than the B and M. This is similar to what is found in
the PNe sample.

The post-AGB morphological distribution is different to that of PNe,
with a
lack of round ones.
There is a strong bias towards the Galactic plane in comparison with the
PNe sample. This may be due to a strong observational bias.

\acknowledgements D.A.G.H. and A.M.  also acknowledges support for this work provided by the Spanish Ministry
of
Science and Innovation (MICINN) under a JdC grant and under grant
AYA$-$2007$-$64748

\bibliography{}
Garc\'{\i}a-Lario, P. et al. 1997, A\&AS, 126, 479\\
Manchado, A. et al. 1989, A\&A, 214, 139\\
Manchado, A. 2003, in Planetary Nebulae: Their Evolution and Role in the
Universe, IAU Symp. 209, eds. S. Kwok, M. Dopita, \& R. Sutherland. PASP,
p.431 \\
Sahai, R. et al. 2007, AJ, 134, 2200\\
Siodmiak, N. et al. 2008, ApJ, 677, 382\\
Suarez, O. et al. 2006, A\&A, 458, 173\\
Szczerba, R. et al. 2007, A\&A, 469, 799\\
Ueta, T. et al. 2000, 528, 861\\

\end{document}